# Web pages search engine based on DNS

Wang Liang[1], Guo Yi-Ping[2], Fang Ming[3]

[1, 3] (Department of Control Science and Control Engineer, Huazhong University of Science and Technology, WuHan, 430074 P.R.China)

[2](Library of Huazhong University of Science and Technology, WuHan 430074 P.R.China)

E-mail: wangliang_f@yahoo.com

Phone: +86-27-87553494

**Abstract:** Search engine is main access to the largest information source in this world, Internet. Now Internet is changing every aspect of our life. Information retrieval service may be its most important services. But for common user, internet search service is still far from our expectation, too many unrelated search results, old information, etc. To solve these problems, a new system, search engine based on DNS is proposed. The original idea, detailed content and implementation of this system all are introduced in this paper.

## 1 Introduction

When designing a search engine we may meet two main bottleneck problems. Because the WWW is a large distributed and dynamic world, search engine can't continue to index close to the entire Web as it grows and changes. So we will meet some serious problems in coverage and recency. According to the statistical data in 1998[1], the update interval of most pages database is almost one month and no a search engine can cover more than 50 percentage pages on Internet. Till now, these data is still available. Although there was no any obvious improvement, there were surely some effective efforts for a better web pages search engine.

Harvest [2] is a representative distributed information retrieval system. Based on Harvest, Cooperate Search Engine (CSE) [3] is developed. These two methods require each web site indexing their web documents and provide interface for search engines. These approaches will reduce the update interval and network traffic, but none of them is widely applied. This is mainly because not all administrators of sites will agree to index their pages for search engines. Reference [4] gave a practical idea to build a distributed search engine. In this paper, Author advised to share the databases of the search engine and introduced a layered architecture to improve the access to data on the Internet. But in this paper, the author didn't give an applied method on how to implement his idea. We develop our new search engine on the foundation of previous works.

## 2 The update of DNS

The original idea of new system could be found in the history of WWW.

When the DNS comes into being, there are only hundreds of web sites, so we can put DNS table in single server. When the number of sites reached level of million and scattered in different place, several DNS can't work efficiently. So DNS developed into a distributed hierarchical system. Now almost all the universities and other big organizations have their own DNS server in local network. All the web sites on Internet are efficiently managed in this system.

But DNS only provide navigation service. We also need searching in WWW. So web search engines appeared. First Yahoo, then Google. For some reasons, all commercial search engines apply centralized architecture. So with the rapid increase of WWW, they also meet the problems that primal DNS meet. There have been billions of pages scattering all over the world. Current search engine have to download them to single database system again and again. The coverage and update interval can't be ensured in this system. Obviously centralized architecture is not appropriate to manage the distributed WWW. Inapposite architecture is the key for most bottleneck problems in current search engine.

Adopting the experience of DNS may be a good selection. Web search engine may also need to apply the distributed hierarchical architecture as DNS. Furthermore, Since DNS can index the name of each site, could they index all the web pages of sites? So idea of "search engine + DNS" appeared.

## 3 Search engine based on DNS

### 3.1 Architecture of new system

As the basic idea of new search engine based on DNS, its architecture is as same as DNS, which is shown in fig1.

There are three layers in this system. The third layer corresponds to some organizations like a university. The second is always the sub Internet of a country. The top layer corresponds to each country. Its three layers strictly correspond to different layers of DNS.

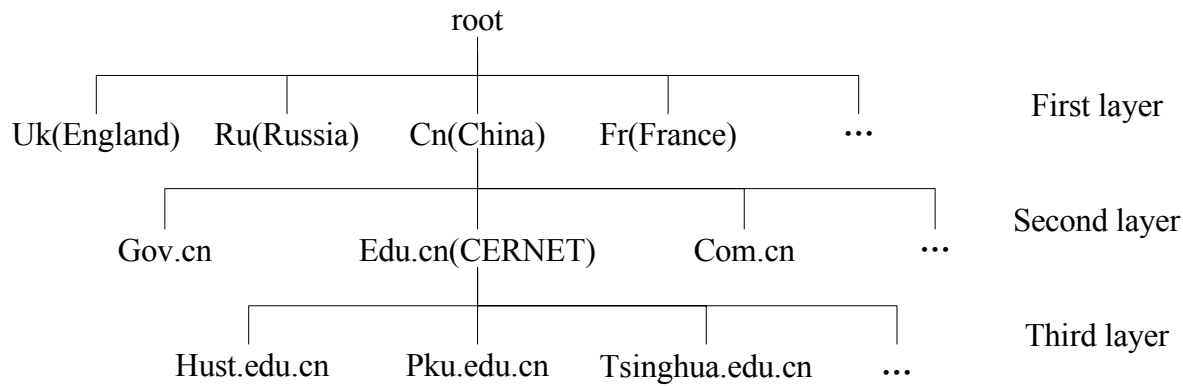

Figure 1  The architecture of the DRIS

In this architecture, we can simply download all the web pages in the bottom layer and send them to the servers in the higher layer. Because all the pages extracting work is done in the different bottom layer, which always corresponds to local network, the data in whole system could be update everyday. So the recency problem could be solved. But in this method, the databases in the top layer will be fairly large. We may have to apply distributed computing or other complex technologies to design such system. At all, building a system that can mirror the whole Internet is almost an impossible mission. We have to find another method to complete it.

**3.2 Three kinds of search system**

According to the architecture, there have been three kinds of information retrieval systems, which are introduced as follows.

1 Centralized search system based on conventional database system. This system has its own data collecting mechanism, and all the data are stored and indexed in a database system. Although many web search engine provide service from thousands servers, their database all are the same and belong to this kind of search system.

2 Search system based metadata harvest system. When we need to integrate different kinds of information resource like video, pdf, web pages in a system, we can harvest the metadata from sub databases and build a union metadata database to provide the combined search function. Normally Metadata is much smaller than data itself. Some systems based on OAI like NSDL just apply this method.

3 Distributed search system. Distributed information retrieval system has no his own actual record database. When receiving a query from a user, it will instantly obtain the records through the search interfaces provided by sub databases. A famous system is the infoBus system in Stanford digital library project [5].

How to select correct information retrieval architecture? There are two main characters to determine the architecture of an information retrieval system, the size and diversity of data source.Normally, with the increasing of size and diversity of data source we will select conventional database system, metadata harvest system and distributed search system respectively. We also apply this principle in the design the new search system based on DNS.

**3.3 Three kinds of search system in three layers**

We will introduce each level's search system from third layer to the top layer.

3.3.1 Third layer: extract and Index the pages

The node in this layer just acts as a normal search engine. Only difference is that the search scope is limited to a third level domain, like a university. Centralized search system based on conventional database system is applied. This search system comprise of three parts: page-fetching mechanism, indexer, and searching interface. The three parts are introduced respectively.

3.3.1.1 Extract the pages

Page-fetching mechanism will download all the pages in a three level domain. For example, "www.hust.edu.cn " is the domain name of our university, so the server under the domain of "hust.edu.cn" like the department of CS's server "cs.hust.edu.cn " can be easily found by referring to the DNS server. Then the spiders can download all sites included in this domain.

The work of spider is organized by single web site. When a spider visits a web server, it will download all the content of this site and stop working when it encounters a URL that links to other site. The content in this "stop URL" is also treated as valuable matter of this site and is downloaded too. This work theory has some differences with conventional search engine, whose spider will freely ramble in WWW and have no stop URL. We have to design complex algorithms to direct spiders. But in new system, its spider just need download all the pages in each site and not consider the intricate relation between each sites.

### 3.3.1.2 Index the pages

Now there is still no a "standard index technology" for the search engines. Normally, the key issue in indexing is the appreciate selection of metadata. In the common sense, a web page is represented by the keywords with its ranking score. We also use this method to index the web pages. By this means, the rank score of keyword will be determined by its position and frequency in the whole document. The other fields such as title, encoding and abstract will also be used to describe a web page. Moreover, we can also use W3C's Ontology [6] model or other advanced technologies to index the web pages.

### 3.3.1.3 Search interface

Providing user interface and processing the search results is main function of search interface. How to rank the pages is the key issue. Now there are two methods for it. First, ranking by the score of keyword. In this method, the score of keywords [7] is calculated by the frequency and the place of this word appearing in the page. When a query word is submitted, all the pages that contain that word will be retrieved and then ranked by the ranking score of corresponding keywords. The other method, Hyperlink Analysis, is applied by some current search engines like Google[8][9]. By this means, the rank score of a page is determined by the number of link in other sites that point to this page.

In this layer, we use the fist method, score of keyword, to rank the results. This is because in this condition the pages are limited to a small area like a university, but the Hyperlink Analysis is more useful in large scale of Internet. So score of keywords is enough.

### 3.3.2 Second layer: Harvest the metadata

This layer will provide the search in all the servers under a second levels domain. Metadata harvest system is applied in this layer. A third layer node like "hust.edu.cn" corresponds to our universities. Most university will have no more than 100 thousand pages, so a centralized search system can work efficiently. But in the second layer node like "edn.cn" will include the web pages in all universities in China. All the pages may exceed 10 million. Conventional search system may not ensure the recency and coverage of its database. So we use metadata harvest system in this layer.

The search engines of this layer have only two parts, web page database and search interface but no its own spider. Its data is downloaded from the databases in the corresponding third layer's nodes. For example, the search engine corresponding to the domain "edn.cn" will obtain its data from the search engine's databases in thousands of universities in China, but not directly from millions of web pages. By this means, the update interval can be much shorter than conventional methods. Only storing the metadata will also make its database fairly small.

A notable problem that should be concerned is the overlap of web pages. In the third layer, when the spider extracts pages from a site, it also gets some pages which don't belong to this site (Stop URL). So when harvesting the metadata, some pages may appear many times. As the download method in the third layer, this number is the number that the other sites direct to this page. According to the theory of Hyperlink Analysis, this number is just the ranking score of this page. Although this score is not calculated in conventional methods, it surely represents the same thing. Obviously Hyperlink Analysis will be more efficient in this scope. It's just the basic work theory. Detailed content like how they cooperate to transfer the metadata will be defined in the standard protocols of this system.

3.3.3 Top layer: a distributed search system

In this layer, storing billions of changeable web pages in single system will meet many bottleneck problems. So the engine in this layer will be a distributed information retrieval system. It has merely one part, search interface, no spider, and no page databases.

There are three issues when designing a distributed search system [10].1 the underlying transport protocol for sending the message (e.g., TCP/IP). 2 a search protocol that specifies the syntax and semantics of message transmitted between client and servers.3 a mechanism that can combine the results from different servers. These problems are detailed as follows.

1 Communication protocol. In this system, SOAP is applied as the basic protocol for communication, because the SOAP has many advantages than HTTP in security, opening, etc.

2 search protocol. Its search protocol is based on Webservice. Webservice applies the SOAP as the fundamental protocol and provides an efficient distributed structure. We refer to the SDLIP [11] and Google's search Webservice to define the format of query and results. By this means, all the search engines in the second layer should provide the standard search Webservice according to a standard protocol. And Search engine in top layer just needs to index all the Webservice in lower layer.

3 combining the results. The key problem for the combination of results is the rank of pages from different servers. In the second layer, the ranking score of one page is represented by its overlap number in databases. In this layer, we also use this theory to calculate the rank score of a document. The score of the same page from different database will be simply added up and a finial ranking list of documents will be produced.

The work theory of this search engine is just like that of Meta search engine, which has no its own pages database, but an index of the search interface of other engines. But the sub search engine based on DNS is arranged strictly and complied with the same protocol, so its performance is much better than any current Meta search engine. The search engine in this layer will provide the search service in a country, which will cover the most of search request on Internet. If you want to search parallel in several countries, you can use the search Webservice in top layer to build a special search engine.

## 3 The advantage of new system

Normally there are three principles to judge a search engine.

1. The coverage of the search engine. The more pages are indexed, the better a search engine is. In theory, the search engine based on DNS can cover almost all the pages on the Internet.
2. The update interval of search engine. All the work of downloading the pages is carried out by many distributed spiders in the third layer, so the content of search engine's databases can be updated everyday.
3. Precise of search results. In fact, all the nodes of search engine based on DNS will provide standard search interface. Many intelligent personal search systems could be designed according your personal information like interests. New search system will be treated as their data source. You will get more precise search results in your personal search engine.

## 4 Implementation of new system

Although search engine based on DNS gives us an excellent and promising solution for the new Internet search system, this can't ensure its establishment. We will meet a common problem: who are willing to build such as system. Finding the request of users is the key for any new system. The distinct architecture of SE based on DNS will ensure it a practical system. In the bottom layer of system, it will build the search engine in the scale of an organization like a university. Now some universities have purchased the commercial search engine for their school network. If a free system appears, most organizations will apply it. Once the systems in bottom layer are completed, we will have the foundation for the higher layer system. All the further work will be directed by IETF.

## 5 Conclusions

Search engine based on DNS has another character. It's a public information retrieval system, but not a commercial search engine. But some companies can design better search system based on this public platform. In fact, almost all the Internet technologies, from IP to E-mail, are all public technology, but the better commercial service could be designed based on these technologies. It may be basic principle for the continued development of Internet.